# The Maximum Energy and Spectra of Cosmic Rays Accelerated in Active Galactic Nuclei

A.V. Uryson

Lebedev Physics Institute of Russian Academy of Sciences, Moscow

e-mail: uryson@sci.lebedev.ru

**Abstract**. We computed the energy spectra of the incident (on an air shower array) ultrahigh-energy ($E > 4 \cdot 10^{19}$ eV) cosmic rays (CRs) that were accelerated in nearby Seyfert nuclei at redshifts $z £ 0.0092$ and in BL Lac objects. These were identified as possible CR sources in our previous works. For our calculations, we took the distribution of these sources over the sky from catalogs of active galactic nuclei. In accordance with the possible particle acceleration mechanisms, the initial CR spectrum was assumed to be monoenergetic for BL Lac objects and a power law for Seyfert nuclei. The CR energy losses in intergalactic space were computed by the Monte Carlo method. We considered the losses through photopion reactions with background radiation and the adiabatic losses. The artificial proton statistic was $10^5$ for each case considered. The maximum energy of the CRs incident on an external air shower (EAS) array was found to be $10^{21}$ eV, irrespective of where they were accelerated. The computed spectra of the particles incident on an EAS array agree with the measurements, which indirectly confirms the adopted acceleration models. At energies $E ³ 10^{20}$ eV, the spectrum of the protons from nearby Seyfert nuclei that reached an EAS array closely matches the spectrum of the particles from BL Lac objects. BL Lac objects are, on average, several hundred Mpc away. Therefore, it is hard to tell whether a blackbody cutoff exists or not by analyzing the shape of the measured spectrum at $E \geq 5 \cdot 10^{19}$ eV.

## 1. INTRODUCTION

At present, the cosmic rays (CRs) with energies $E > 4 \cdot 10^{19}$ eV are generally believed to be extragalactic in origin, but their sources have not been firmly established. Various astrophysical objects, cosmological defects, decaying superheavy primordial particles of cold dark matter, and gamma-ray bursts are considered in the literature as their possible sources (see the review [1] and references therein). In the first case, the CR sources can be identified if the CR arrival directions are known and if the particles are assumed to propagate in the intergalactic magnetic fields almost rectilinearly. We directly identified the possible sources of ultrahigh-energy CRs previously [2-4] and found them to be Seyfert nuclei at redshifts $z £ 0.0092$ and BL Lac objects. BL Lac objects were also identified with possible CR sources in [5]. Authors of [6] and [7, 8] suggested a particle acceleration mechanism in BL Lac objects and moderately active galactic nuclei, respectively. According to [6], CRs can be accelerated in BL Lac objects up to $10^{27}Z$ eV, where $Z$ is the particle charge, and, if there are energy losses in the sources, up to $10^{21}Z$ eV. In Seyfert nuclei, particles can be accelerated up to an energy of $8 \cdot 10^{20}$ eV [7, 8]. In intergalactic space, particles interact with background radiation; as a result, they inevitably lose their energy [9, 10]. Particles of different energies traverse different distances without significant energy losses. These distances for ultra high energy CRs were estimated in [11, 12]. The redshifts $z £ 0.0092$ of Seyfert nuclei correspond to distances up to 40 Mpc (for the Hubble constant $H = 75$ km s$^{-1}$ Mpc$^{-1}$), in agreement with the results [11, 12]. The BL Lac objects identified as possible CR sources are far, up ~1000 Mpc [13], away. Therefore, the question arises as to whether the particles with an energy of $3 \cdot 10^{20}$ eV (the maximum recorded CR energy [14]) accelerated in BL Lac objects can reach an EAS array. In this paper, we computed the energies of the particles from BL Lac objects that reached an EAS array and the energy spectra of the incident (on an EAS array) CRs that escaped from active galactic nuclei (AGNs) with power-law and monoenergetic spectra. We compared the computed and measured spectra. For our calculations, we took the distribution of AGNs from the catalog [13].

## 2. DESCRIPTION OF THE MODEL

According to the model [6], particles in BL Lac objects are accelerated in the electric field induced near a supermassive black hole with a mass of ~$10^9 M_O$, where $M_O$ is the mass of Sun. Particles are accelerated in this field up to an energy of $10^{27}Z$ eV; the particle energy can decrease through curvature radiation to $10^{21}Z$ eV. Based on this acceleration mechanism, we assume in our calculations that the initial spectrum of the protons accelerated in BL Lac objects is monoenergetic with an initial energy of $10^{27}$ and $10^{21}$ eV. Since particles in Seyfert nuclei maybe accelerated at shock fronts [7], we assume that the initial spectrum of the particles from them is a power law ($\sim E^{-c}$) with a spectral index of $? = 2.6$ and $3.0$. Particles in Seyfert nuclei can be accelerated up to an energy of $8 \cdot 10^{20}$ eV.

The composition of the CRs with energies $E \approx 4 \cdot 10^{19} - 3 \cdot 10^{20}$ eV is not yet completely known. In accordance with the data [15], we assume that the CRs with energies as high as $10^{21}$ eV are particles rather than gamma-ray photons.

The propagation of CRs in intergalactic space was considered under the following assumptions. The nuclei disintegrate into nucleons through their interactions with background radiation, traveling no more than 100 Mpc from their source [12, 16]. Therefore, if the CR sources are much farther than 100 Mpc, then, for simplicity, we may assume that the nuclei completely fragment near the source and consider only the propagation of protons in intergalactic space. Since the overwhelming majority of BL Lac objects are $R > 400$ Mpc away [13], this assumption is justified for the CRs emitted by BL Lac objects. For simplicity, we assume that only protons propagate from Seyfert nuclei as well.

We computed the CR energy losses in intergalactic space under the following assumptions. Protons interact with relic and infrared photons. Protons with energies $E > 4 \cdot 10^{19}$ eV lose their energy mainly through the photopion reactions $p + ? \rightarrow N + p$; the losses through the electron–positron pair production are negligible [17, 18]. The density spectrum for relic photons with energy $e$ is described by the Planckian distribution

$$n(e)de = e^2 de / (p^2 h^3 / (2\pi)^3 c^3 (\exp(e/kT)-1)) \qquad (1)$$

with the temperature $T = 2.7$ K, the mean photon energy is $\langle e \rangle \approx 6 \cdot 10^{-4}$ eV, and their mean density is $\langle n_0 \rangle \approx 400$ cm$^{-3}$. For the photons of the high energy tail in the Planckian distribution, the mean energy is $\langle e_t \rangle \approx 1 \cdot 10^{-3}$ eV and the mean density is $\langle n_t \rangle \approx 42$ cm$^{-3}$.

The energy range of the infrared radiation is $2 \cdot 10^{-3} - 0.8$ eV; at present, there are no detailed spectral measurements. We assumed that the infrared radiation spectrum is described by the numerical expression [12, 16]

$$n(e) = 7 \cdot 10^{-5} e^{-2.5} \text{ cm}^{-3} \text{ eV}^{-1}, \qquad (2)$$

the mean energy of the infrared photons is $\langle e_{IR} \rangle \approx 5.4 \cdot 10^{-3}$ eV, and their mean density is $\langle n_{IR} \rangle \approx 2.28$ cm$^{-3}$.

The photopion reactions are threshold ones. The threshold energy is $e_{th}* \approx 145$ MeV, where $e*$ is the photon energy in the proton frame of reference; the threshold inelasticity coefficient is $K_{th} \approx 0.126$ [11]. The cross section $s$ and the inelasticity coefficient $K$ of the photoprocesses depend on the energy $e*$. The dependences $s(e*)$ and $K(e*)$ were taken from [11, 19]. The values of $s$ and $K$ used in our calculations are given in the table.

In addition to the photopion reactions, protons lose their energy through the expansion of the Universe. The adiabatic losses of a proton that propagates with an initial energy $E$ from a point with a redshift $z$ to a point with $z = 0$ are

$-dE/dt = H(1 + z)^{3/2} E.$  (3)

The cosmological evolution of the Universe was taken into account in the CR propagation. We used the Einstein–de Sitter model with $\Omega = 1$, in which the time and the redshift are related by

$t = 2/3 H^{-1}(1 + z)^{-3/2};$  (4)

the distance to an object at redshift $z$ is

$r = 2/3 c H^{-1}[1-(1 + z)^{-3/2}]$ Mpc.  (5)

At the epoch with a redshift $z$, the relic photon density and energy were, respectively, a factor of $(1 + z)^3$ and $(1 + z)$ higher than those at $z = 0$ [18].

We assumed that the particles propagate in the intergalactic magnetic fields almost rectilinearly. The sources of ultrahigh-energy CRs, BL Lac objects and Seyfert nuclei at $z \leq 0.0092$, were assumed to be distributed in redshift in accordance with the catalog [13]. The $z$ distributions of these objects at declinations $d \geq 15^0$ are shown in Figs. 1 and 2.

## 3. CALCULATIONS

The calculations were performed as follows. First, we generated the redshift $z_0$ of a source by the Monte Carlo method in accordance with the distributions shown in Figs. 1 and 2. Subsequently, we calculated the distance to the source. Since the energy losses of the CRs depend on the distances that they traverse in intergalactic space, we determined them by two methods to reach reliable conclusions. The first method uses formula (5). The second method assumes that $r = czH^{-1}$ (Mpc) for $z < 0.4$ [20] and uses formula (5) for higher $z$. The calculations were performed with $H = 75$ and $100$ km s$^{-1}$ Mpc$^{-1}$. Next, we randomly generated the proton energy $E$ and the angle ? in the laboratory frame and determined the photon energy in the proton frame

$e^* = ?e(1-ß \cos ?),$  (6)

where ? is the Lorentz factor of the proton, and $ß = (1-1/?^2)^{1/2}$. If $e^* < e^*_{th}$, then the proton interacted with photons of the high-energy tail in the Planckian distribution. If, alternatively, the photon energy $e^*$ was also below its threshold value in this case, then the proton interacted with infrared photons. The cross section $s$ and the inelasticity coefficient $K$ for this interaction were determined from the value of $e^*$. Subsequently, we calculated the proton mean free path $<?> = (<n>s)^{-1}$, where $<n>=<n_o>$, $<n_t>$ or $<n_{IR}>$, depending on which photon the proton interacted with. Next, we generated the proton mean free path $L$ by the Monte Carlo method and calculated the redshift $z_1$ of the proton after it traversed the distance $L$. At the point with $z_1$, the proton energy decreased due to its interaction with the photon by $(\Delta E)_{ph} = EK$. The decrease in energy due to the adiabatic losses is

$(\Delta E)_{ad} = E(z_0 - z_1)/(1 + z_0).$  (7)

This procedure was then repeated. In our calculations of the adiabatic losses at the point with redshift $z_2$, the point of the preceding interaction with redshift $z_1$ in formula (7) was taken in place of the point with $z_0$, the point with $z_2$ was taken in place of the point with $z_1$, an so on. The procedure ended if the proton reached the Earth (the point with $z_i = 0$) or if its energy decreased to $E < 4 \cdot 10^{19}$ eV.

## 4. RESULTS

### 4.1. The Maximum Particle Energy in a Source

Authors of [21, 22] theoretically estimated the maximum CR energy in sources to be $\sim 10^{21}$ eV. The initial proton energy in BL Lac objects without and with the inclusion of curvature losses in the source is, respectively, $10^{27}$ and $10^{21}$ eV [6]. These estimates can be easily compared with the CR data by calculating the mean energies of the protons with initial energies of $10^{27}$ and $10^{21}$ eV incident on an EAS array. We computed the energies of the incident (on an EAS array) protons from BL Lac objects that were distributed as in Fig. 2 and that had the above initial energies using the Monte Carlo method. In each case, the artificial proton statistic was $10^4$. The mean proton energies on Earth were found to be $10^{24}$ and $6 \cdot 10^{19}$ eV, respectively. The first value is in conflict with the experimental data (recall that based on the possible CR acceleration mechanism [6] (Kardashev 1995), we assumed the initial spectrum in BL Lac objects to be monoenergetic), the calculation with an initial energy of $10^{21}$ eV is consistent with the measurements, in agreement with its theoretical estimate [21, 22]. This value is close to the maximum energy, $8 \cdot 10^{20}$ eV, of the particles emitted by Seyfert nuclei [7, 8]. For the subsequent analysis, let us consider the proton spectra on Earth.

### 4.2. The Spectra of the Protons Incident on an EAS Array

The measured CR spectrum at $E > 4 \cdot 10^{19}$ eV exhibits a flat component and a bump that are probably attributable to the CR energy losses in intergalactic space: these losses lead to the "transfer" of particles to the range of lower energies provided that the energy losses decrease with decreasing energy [23, 24]. Authors of [25-27] computed the CR spectrum and analyzed its shape. The closer the source, the higher the energy at which a bump appears in the spectrum. We analyzed the shape of the measured spectrum in the energy range $10^{18}$–$10^{20}$ eV previously [28]. Since the energies of the particles that trigger air showers are measured by different methods, the CR spectra measured on different EAS arrays differ in intensity. The combined spectra normalized using measurements on a particular EAS array are given in the literature. Here, we compare the computed spectra with published measurements.

The differential ultrahigh-energy CR spectra measured on different EAS arrays and normalized in different ways are shown in Fig. 3: the spectra from [1] normalized using AGASA data are shown in Figs. 3a and 3c; the spectra obtained on the same arrays and on the HiRes array and normalized using the Fly's Eye array are shown in Fig. 3b [29]. The computed spectra normalized using measurements are shown in the same figures. The artificial proton statistic is $10^5$ for each curve. Let us first consider the spectra in Fig. 3a. The large measurement errors make it difficult to compare the computed curves with the experimental data. However, two models are clearly inconsistent with the measurements: the models with a monoenergetic initial spectrum in Seyfert nuclei and with a power-law initial spectrum in BL Lac objects. The models with a monoenergetic initial spectrum in BL Lac objects and with a power-law initial spectrum in Seyfert nuclei are suitable for describing the data, but the initial spectral index, 3.0 or 2.6, is difficult to determine due to the large errors. This is consistent with the possible particle acceleration conditions in these sources.

The measured spectrum in Fig. 3b agrees with these curves, except for the HiRes data points at $E < 10^{20}$ eV. The HiRes data are best described by the model with a power-law spectrum in BL Lac objects at $? = 2.0$ and by the model with a power-law spectrum in Seyfert nuclei at $? = 3$. Figure 3c shows the same data as in Fig. 3a, but the curves were computed for $H = 100$ km s$^{-1}$ Mpc$^{-1}$ where the distances were determined by the second method (see the section "Calculations"). In this case, the measurements are also described by the models with a monoenergetic initial spectrum in BL Lac objects and with a power-law initial spectrum in Seyfert nuclei.

Thus, the data from different EAS arrays are described by the model in which the CR sources are nearby Seyfert nuclei with a power-law initial spectrum. The model in which the CR sources are BL Lac objects with a monoenergetic initial spectrum is also suitable for describing the data, except for the HiRes data.

At $E > 10^{20}$ eV, the spectra computed in the models with a monoenergetic spectrum in BL Lac objects and with a power-law spectrum in nearby Seyfert nuclei satisfactorily describe the measurements and are very similar. The computed CR spectra in these models will differ greatly if 2% of the BL Lac objects at redshifts $z < 0.1$ are assumed to be at the distance with $z = 0.01$ (according to the catalog [13], the minimum redshift for BL Lac objects is $z = 0.02$). The spectra computed under this assumption are shown in Fig. 3c.

It follows from the above analysis that the models for both far and nearby sources account for the measured CR spectrum at energies $E > 4 \cdot 10^{19}$ eV. Therefore, analyzing the spectrum in this energy range, we currently cannot determine whether it has a blackbody cutoff. In addition, AGASA, HiRes, Fly's Eye, Haverah Park, and Yakutsk data indirectly confirm our model of particle acceleration in nearby sources. Data from the arrays, except for HiRes, also confirm the model of particle acceleration in BL Lac objects.

At energies below $10^{19}$ eV, the spectrum may be shaped by particles from distant sources [18, 27]. According to the currently available data [13], the total number of Seyfert nuclei and BL Lac objects is several thousand and several hundred, respectively.

### 4.3. Estimates of the CR Luminosity for Sources

We estimated the CR luminosity of Seyfert nuclei previously [3, 4]: $(L_{CR})_S \approx 10^{40}$ erg s$^{-1}$ for $? = 3$ in the power-law initial CR spectrum and $(L_{CR})_S \approx 10^{42}$ erg s$^{-1}$ for $? = 3.1$. The actual power spent on the CR acceleration in a source is a factor of ~300 higher due to the curvature radiation of particles in the source.

Let us estimate the observed CR luminosity for BL Lac objects $(L_{CR})_{BL\ LAC}$:

$$L_{CR})_{BL\ LAC} = U_{CR}/(NT), \qquad (8)$$

where $U_{CR}$ is the total energy of the CRs emitted by BL Lac objects, $N$ is the total number of BL Lac objects, and $T$ is the CR lifetime. We can determine $U_{CR}$ from the energy balance equation:

$$U_{CR} = (U_{CR})_{measured} + (U_{CR})_{lost}, \qquad (9)$$

where $(U_{CR})_{measured}$ is the energy of the CRs that reached the EAS array, and $(U_{CR})_{lost}$ is the CR energy that was lost during the CR propagation from the source to the EAS array. The initial CR energy in the source is $E_0 = 10^{21}$ eV; the bulk of the CRs on the EAS array have an energy of $E = 5 \cdot 10^{19}$ eV. Assuming that

$$(U_{CR})_{measured}/U_{CR} \approx E/E_0 \approx 0.05, \qquad (10)$$

we obtain

$$U_{CR} \approx 20(U_{CR})_{measured}. \qquad (11)$$

We define $(U_{CR})_{measured}$ as

$$(U_{CR})_{measured} = \int_E I(E) E dE 4\pi/cV, \qquad (12)$$

where $I(E)$ is the computed intensity of the CRs from BL Lac objects, and $V$ is the CR-filled volume. The integral in (12) is equal to ~4 eV cm$^{-2}$ s$^{-1}$ sr$^{-1}$. Most of the BL Lac objects have redshifts $z \leq 0.35$ (see Fig. 2); i.e., they are $r \leq 1000$ Mpc away. The CRs emitted by these sources reach an EAS array in a time $T \leq 2 \cdot 10^{17}$ s. Assuming that the CRs fill a sphere with a radius $r \approx 1000$ Mpc and reach an EAS array in a time $T \approx 2 \cdot 10^{17}$ s, we find that the total power of the sources is $U_{CR}/T \approx 2 \cdot 10^{44}$ erg s$^{-1}$. The number of sources at redshifts z≤0.35 is $N \approx 100$ [13]. Therefore, the CR luminosity of a single source is $(L_{CR})_{BL\ LAC} \approx 2 \cdot 10^{42}$ erg s$^{-1}$. (The number of BL Lac objects may be much larger; the luminosity $(L_{CR})_{BL\ LAC}$ is then lower than the value obtained above.)

The power spent on the CR acceleration in a source is higher than its observed value, $2 \cdot 10^{48}$ (erg s$^{-1}$), because we assumed in our estimates that the initial particle energy is $10^{21}$ eV, while these particles are accelerated in the source up to $10^{27}$ eV. According to the model [6], CRs emerge from a source with an energy of $10^{21}$ eV due to the curvature losses, and the bulk of the energy is spent on gamma-ray radiation.

## 5. DISCUSSION

The maximum CR energy is $10^{21}$ eV, irrespective of where they were accelerated, in Seyfert nuclei or in BL Lac objects. This energy is close to the values obtained in the models [30-32]: ~$10^{21}$ eV for the CRs accelerated in an accretion disk around a black hole with a mass of ~$10^7 M_O$, ~$3 \cdot 10^{21}$ eV if the particles are produced in the decays of metastable superheavy particles of cold dark matter, and ~ $10^{21}$ eV if the CRs are accelerated in gamma-ray bursts. The maximum accelerated particle energy of $10^{21}$ eV was also obtained in [21, 22]. The value of ~$10^{20}$ eV predicted in the model [33] appears to be incorrect. In this model, CRs are accelerated in the galactic magnetic fields by a surfatron mechanism. However, particles at energies of $10^{19}$ eV are not confined by the Galactic magnetic fields, and their capture by suitable (for the subsequent surfatron acceleration) shock waves probably becomes impossible.

## 6. CONCLUSIONS

The observed CR luminosity for Seyfert nuclei is $(L_{CR})_S \approx 10^{40}$ erg s$^{-1}$ if ? = 3 in the power-law initial CR spectrum; for BL Lac objects, the observed CR luminosity is $(L_{CR})_{BL\ LAC} \approx 2 \cdot 10^{42}$ erg s$^{-1}$. The power spent on the CR acceleration in sources is much higher: $3 \cdot 10^{42}$ erg s$^{-1}$ for Seyfert nuclei and $2 \cdot 10^{48}$ erg s$^{-1}$ for BL Lac objects. The bulk of the energy lost in the source is spent on gamma-ray radiation.

The model in which CRs are accelerated with a power-law initial spectrum in nearby Seyfert nuclei satisfactorily describes the AGASA, HiRes, Fly's Eye, Haverah Park, and Yakutsk measurements. Data from the EAS arrays, except for HiRes, also confirm the model in which CRs are accelerated with a monoenergetic initial spectrum in BL Lac objects. The maximum CR energy is $10^{21}$ eV.

The models for both far and nearby sources satisfactorily describe the measured CR spectrum. Consequently, there is no paradox in the fact that far BL Lac objects have been identified as possible CR sources. In addition, in the model at $E > 4 \cdot 10^{19}$ eV, the spectrum of the particles arrived from nearby Seyfert nuclei is similar to the particle spectrum from far BL Lac objects. Therefore, analyzing the spectrum in this energy range, we cannot determine whether it has a blackbody cutoff.

It follows from these results that the ultrahigh-energy CR spectrum can be an additional test for the models of sources used here: whether the acceleration conditions in them are indeed such that the initial spectrum is monoenergetic in BL Lac objects and a power law in Seyfert nuclei. The ultrahigh-energy CR spectrum will be determined with a higher energy resolution and a larger statistic in HiRes, Auger, and Telescope Array measurements as well as in satellite measurements [1, 34].

**ACKNOWLEDGMENTS**

I wish to thank N.S. Kardashev for a discussion of the AGN models and V.A. Tsarev for a discussion of the ultrahigh-energy CR data. I am grateful to the referees for remarks.

The table. Cross sections $s$ and inelasticity cofficients $K$ from [11, 19] used in our calculations.

| $\varepsilon^*$, GeV | .145 | .2 | .3 | .33 | .4 | .5 | .6 | .7 | .8 | .9 | 1. | 1.4 | 2. | 3. | 4. | 5. |
|---|---|---|---|---|---|---|---|---|---|---|---|---|---|---|---|---|
| K |  | .13 | .145 | .2 | .215 | .23 | .25 | .3 | .33 | .35 | .38 | .4 | .46 | .5 | .5 | .5 | .5 |
| $\sigma$, mbarn | .05 | .08 | .4 | .43 | .35 | .23 | .22 | .215 | .2 | .19 | .17 | .125 | .14 | .095 | .073 | .07 |

**Figures Captions**

**Fig. 1.** Redshift distribution of nearby ($z \leq 0.0092$) Seyfert galactic nuclei normalized to the total number of objects.

**Fig. 2.** Redshift distribution of BL Lac objects normalized to the total number of objects.

**Fig. 3.**

**(a)** Differential CR energy spectra as measured on different EAS arrays (Haverah Park, Fly's Eye, AGASA, and Yakutsk) from [1]Nagano and Watson (2000). The curves represent the spectra computed for $H = 75$ km s$^{-1}$ Mpc$^{-1}$ where the distances were determined by the first method; the solid lines indicate the spectra of the CRs arrived from BL Lac objects: a power-lawinitial spectrum in the sources with $\chi = 3.0$ (1), a power-lawinitial spectrum with $\chi = 2.6$ (2), and a monoenergetic initial spectrum (3); the dashed lines indicate the spectra of the CRs arrived from Seyfert galactic nuclei: a power-law initial spectrum with $\chi = 3.0$ (4), a power-law initial spectrum with $\chi = 2.6$ (5), and a monoenergetic initial spectrum (6).

**(b)** The differential CR energy spectra as measured on different EAS arrays (Fly's Eye, HiRes, AGASA, and Yakutsk) from [29] Bahcall and Waxman (2003). The spectra were computed in the same way as those in Fig. 3?. The solid lines indicate the spectra of the CRs arrived from BL Lac objects: a power-law initial spectrum in the source with $\chi = 2.6$ (1), a power-law initial spectrum with $\chi = 2.0$ (2), and a monoenergetic initial spectrum (3); the dashed lines indicate the spectra of the CRs arrived from Seyfert galactic nuclei: a power-law initial spectrum with $\chi = 3.0$ (4) and a power-law initial spectrum with $\chi = 2.6$ (5).

**(c)** The same as Fig. 3a, but the spectra were computed for $H = 100$ km s$^{-1}$ Mpc$^{-1}$ where the distances were determined by the second method. The letter *a* mark the spectra of the CRs arrived from BL Lac objects for $z_{min} = 0.01$; in the remaining cases, the spectra from BL Lac objects were computed for $z_{min} = 0.02$.

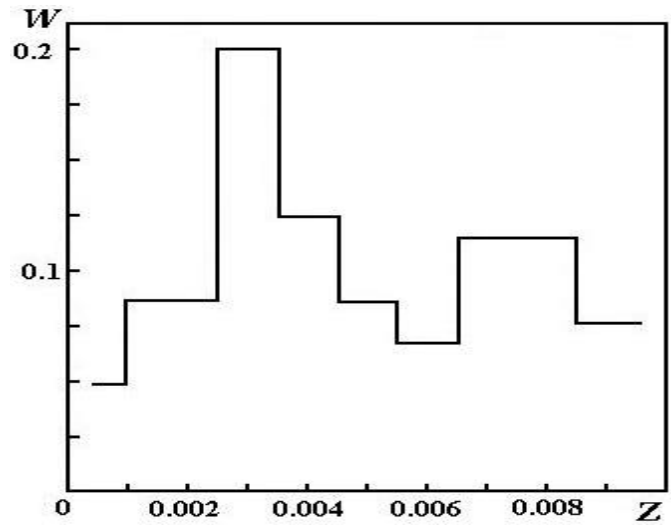

**Fig.1.**

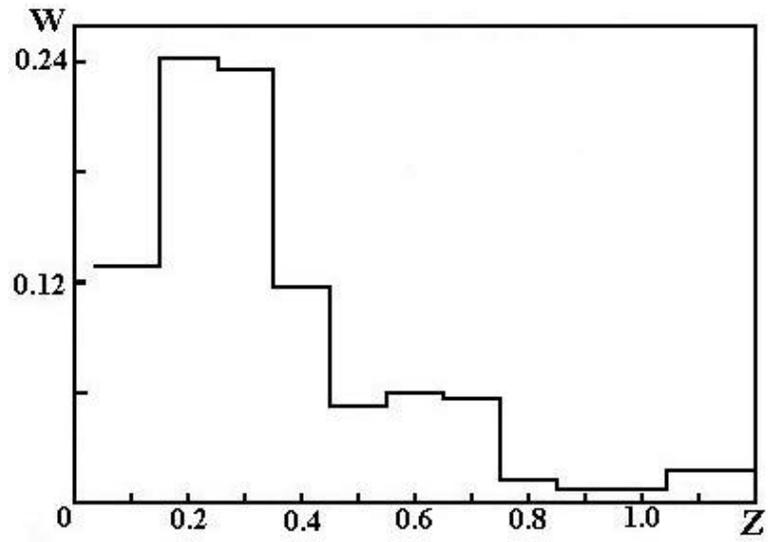

**Fig.2.**

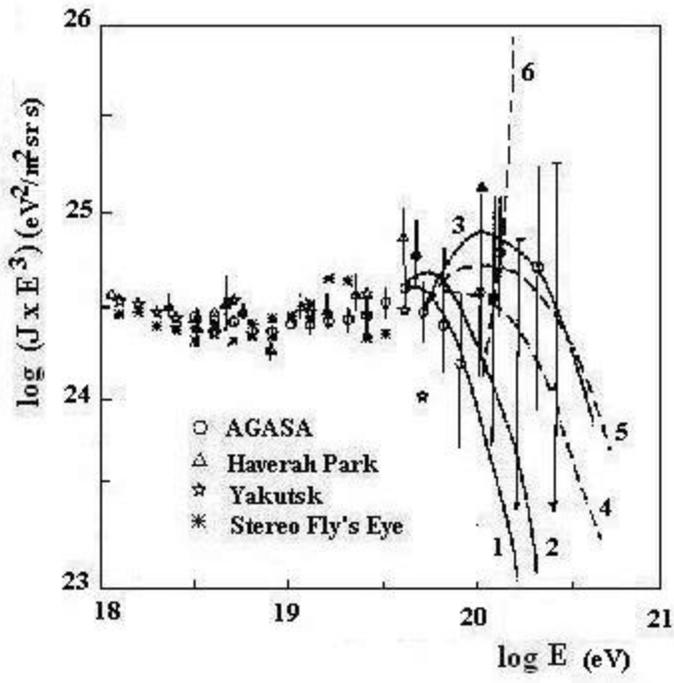

**Fig.3a.**

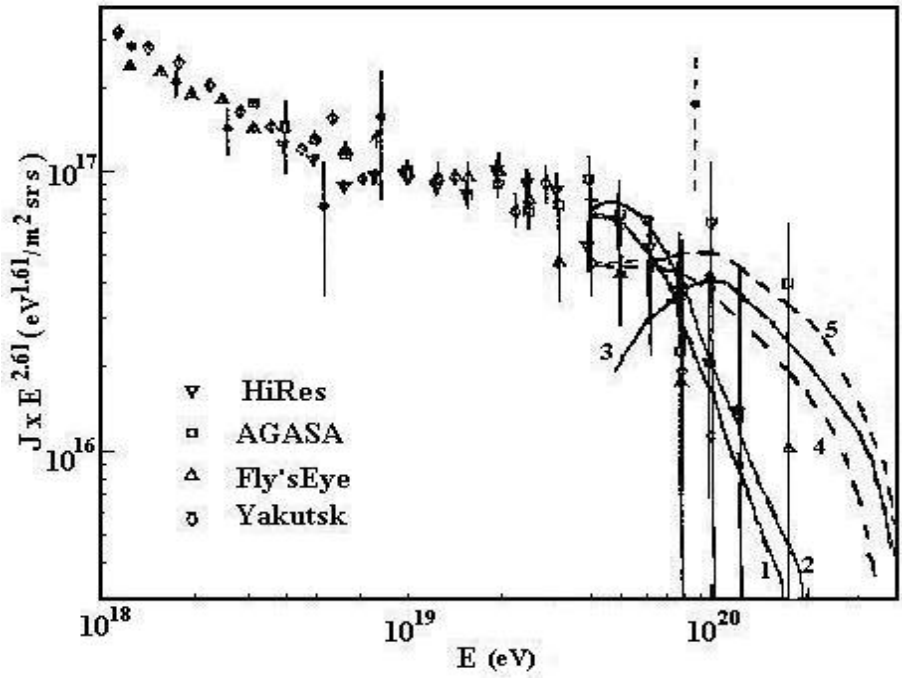

**Fig3b.**

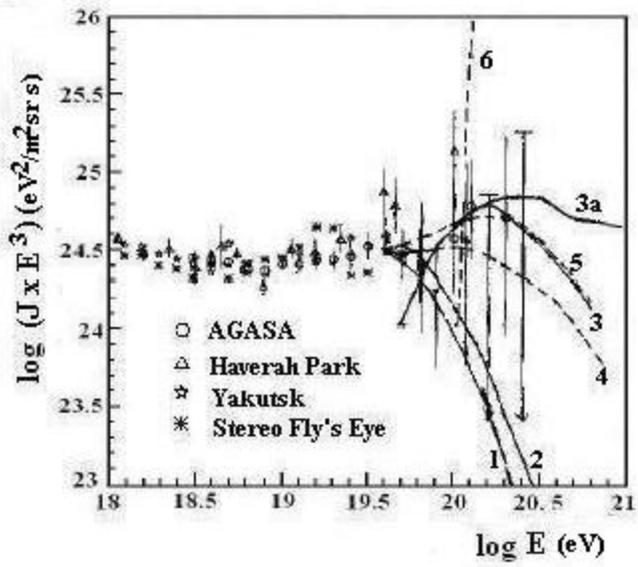

**Fig.3c.**